\documentclass[prl,twocolumn,showpacs,preprintnumbers,amsmath,amssymb,superscriptaddress,nofootinbib]{revtex4}

\usepackage{graphics}
\usepackage{graphicx}
\usepackage{epsfig}
\usepackage{epsf}
\usepackage{bm}
\usepackage{subfigure}
\usepackage{amsmath}

\usepackage{amssymb}

\begin{document}




\title{Narrowband Emission in Thomson Sources Operating in the 
High-Field Regime}


\author{Bal{\v s}a Terzi{\'c}}
\affiliation{Jefferson Lab, Newport News, Virginia 23606, USA}
\affiliation{Center for Accelerator Studies, Old Dominion University, Norfolk,
Virginia 23539, USA}

\author{Kirsten Deitrick}
\affiliation{Center for Accelerator Studies, Old Dominion University, Norfolk,
Virginia 23539, USA}

\author{Alicia S.~Hofler}
\affiliation{Jefferson Lab, Newport News, Virginia 23606, USA}

\author{Geoffrey A.~Krafft}
\affiliation{Jefferson Lab, Newport News, Virginia 23606, USA}
\affiliation{Center for Accelerator Studies, Old Dominion University, Norfolk,
Virginia 23539, USA}

\begin{abstract}
We present a novel and quite general analysis of the interaction of a 
high-field chirped laser pulse and a relativistic electron, in which 
exquisite control of the spectral brilliance of the upshifted 
Thomson-scattered photon is shown to be possible. Normally, when Thomson
scattering occurs at high field strengths, there is ponderomotive line 
broadening in the scattered radiation.  This effect makes the bandwidth 
too large for some applications and reduces the spectral brilliance. 
We show that such broadening can be corrected and eliminated 
by suitable frequency modulation of the incident laser pulse. Further, 
we suggest a practical realization of this compensation idea in terms of 
a chirped-beam driven free electron laser oscillator configuration 
and show that significant compensation can occur, even with the 
imperfect matching to be expected in these conditions.
\end{abstract}


\maketitle

Sources of electromagnetic radiation relying upon Thomson scattering
are increasingly being applied in fundamental physics research
\cite{krafftreview}, and compact accelerator-based sources specifically
designed for potential user facilities have been built \cite{ruth}.
One remarkable feature of the radiation 
emerging from such sources, compared to bremsstrahlung sources, is the 
narrowband nature of the radiation produced. For example, applications to 
X-ray structure determination \cite{r2}, dark-field imaging \cite{r3,r3r}, 
phase contrast imaging \cite{r4}, and computed tomography \cite{r5} have 
been demonstrated experimentally and take full advantage of the 
narrow bandwidth of the Thomson source.

Given that narrow bandwidth is desired, it is important to know and understand 
the sources of bandwidth of the scattered radiation and the limitations 
imposed on the performance of Thomson sources. For applications 
where the normalized vector potential of the incident
laser pulse is much less than one (the low-field regime), 
the line width of the radiation from a scattering event reproduces the 
line width of the incident laser pulse. Unfortunately, when the normalized
vector potential increases, 
as is desired for stronger sources, a detuning red-shift arises during 
the scattering events that tends to spread out the spectrum
\cite{Brau,krafft,Gao}. Physically, the scattering electron slows down, 
by a varying amount, as the incident pulse is traversed.

In a recent paper, Ghebregziabher, Shadwick, and Umstadter (GSU) observed
that frequency modulation (FM), or ``chirping'', of the scattering laser pulse
can compensate for such ponderomotive line broadening, and suggested a form
for this modulation \cite{gheb}. Motivated by their 
observation, we present the exact analytic solution for optimal FM, 
recovering the low-field linewidth even in the high-field regime. 
The narrowing of the scattered pulse is Fourier-limited only by the 
duration of the incident pulse.

\begin{figure*}
\begin{center}
\includegraphics[width=6.75in]{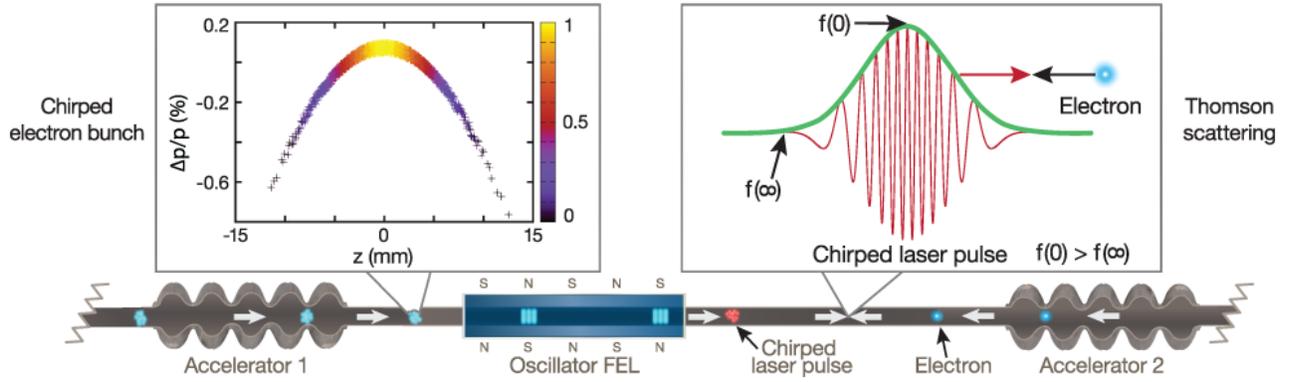}
\caption{\small{
Thomson scattering for a chirped laser pulse.
Chirping the electron bunch before it passes through an oscillator FEL
produces a chirped laser pulse. Thomson scattering occurs as 
the laser pulse collides with an electron. Narrowband emission of the 
resulting backscattered radiation is recovered through frequency
chirping.}}
\label{fig:dist}
\end{center}
\end{figure*}

The essence of laser pulse chirping is analogous to free electron laser (FEL)
undulator tapering \cite{kmr,le11,nel,chris,orz}. In tapering, as 
deceleration occurs due to the FEL emission, the field strength is 
adjusted to preserve the same 
FEL emission frequency. To apply this idea to Thomson sources, 
the field strength dependence of the frequency shift must be 
countered by modulating the incident laser pulse. We 
show below how such modulation is straightforwardly accomplished in 
FEL oscillator lasers using the natural energy phase space curvature 
implicit in a bunch placed on the crest of an RF accelerating wave 
(Fig.~\ref{fig:dist}).

In this Letter, we derive a prescription for calculating the choice of 
the FM that compensates the frequency spreading and recovers the initial 
spectral width. Even at high field strength, one can by proper choice of FM 
reduce the spectral width to the Fourier limit provided by the laser pulse
width. Strictly speaking, the calculations in this Letter apply to Compton 
scattering only in the Thomson limit \cite{krafftreview}, i.e., 
when electron recoil may be neglected. This approximation is valid for Compton
sources of X-rays originating from electron beams up to 100 MeV beam energy.

We report on calculations completed using the formalism developed in 
Ref.~\cite{krafft}, which derives far-field spectral distribution of photons 
Thomson-scattered by a single electron. The incident laser pulse is
described by a plane wave. This treatment is fully relativistic and includes 
the classical electron motion without approximation. We assume a 
linearly polarized incident plane wave described by a single component for 
the normalized vector potential 
${\tilde{A}(\xi})=eA(\xi)/mc=a(\xi)\cos(2\pi\xi f(\xi)/\lambda)$, where 
$a(\xi)$ describes the envelope of the oscillation, $\xi=z+ct$ is the 
coordinate giving distance along the laser pulse, $f(\xi)$ specifies the 
laser FM, and $\lambda$ is a convenient normalizing wavelength for the 
incident plane wave. In previous literature $f(\xi)=1$, but following 
Ref.~\cite{gheb} we allow the possibility of laser chirping and let $f$ 
vary throughout the pulse. Without loss of generality and for convenience 
of presentation, we require $f(0)=1$ for the peak amplitude centered at 
$\xi=0$. Under this convention the calculated spectrum in the forward 
direction has a maximum near the frequency
$\omega_{max}\approx 2\pi c(1+\beta)^2\gamma^2/[\lambda(1+a^2(0)/2)]$,
where $\beta$ and $\gamma$ are the usual relativistic factors for the 
scattering electron. We follow the practice in Ref.~\cite{krafft} and normalize 
frequencies by $\omega_0=(1+\beta)^2\gamma^22\pi c/\lambda$. We report 
calculated spectra normalized by 
$(d^2E/d\omega d\Omega)_n=(1+\beta)^2\gamma r_eE_{beam}/c$ where 
$r_e$ is the classical electron radius and $E_{beam}$ is the total 
relativistic energy in a single electron. These spectra include both 
positive and negative frequency contributions in a single positive 
frequency spectrum. This normalization reflects the main beam energy 
dependence of the spectrum.

Because the longitudinal velocity of the electron changes and there is 
relativistic red-shifting during the laser pulse as the electron traverses 
it, the emitted spectrum of scattered radiation for a constant wavelength 
incident laser pulse is spread out beyond the spectral width 
of the incident pulse. Within each harmonic $n$ in the emitted spectrum 
of scattered radiation, subsidiary peaks are featured whose number $N_{\tau}$
is proportional to the field strength squared and the temporal duration of 
the pulse $T$. We use the Eq.~(31) of \cite{Brau} and a stationary
phase argument \cite{Brau,ck} to derive the exact relationship 
\begin{equation}
N_{\tau} = (2n-1) {c\over\lambda} T \int_0^\infty a^2({\bar \xi})d 
{\bar \xi}, 
\end{equation}
where ${\bar \xi}\equiv {\xi/(\sqrt{2}\sigma)}$, with $\sigma$ the pulse length.
For a gaussian envelope $a(\xi)=a_0 \exp[-\xi^2/(2\sigma^2)]$ and 
$n=1$, $N_{\tau}=\sqrt{2\pi}cTa^2(0)/4\lambda$, 
which agrees well with the empirical formula 
$N_{\tau} \approx 0.24 T [{\rm fs}] a^2(0)$
from Ref.~\cite{Heinzl}.
While their empirical formula applies only to pulses with 
gaussian envelopes at $\lambda=800$ nm, ours is exact and valid for 
arbitrary envelope functions and wavelengths.

\begin{figure}
\begin{center}
\subfigure{
\includegraphics[width=3.375in]{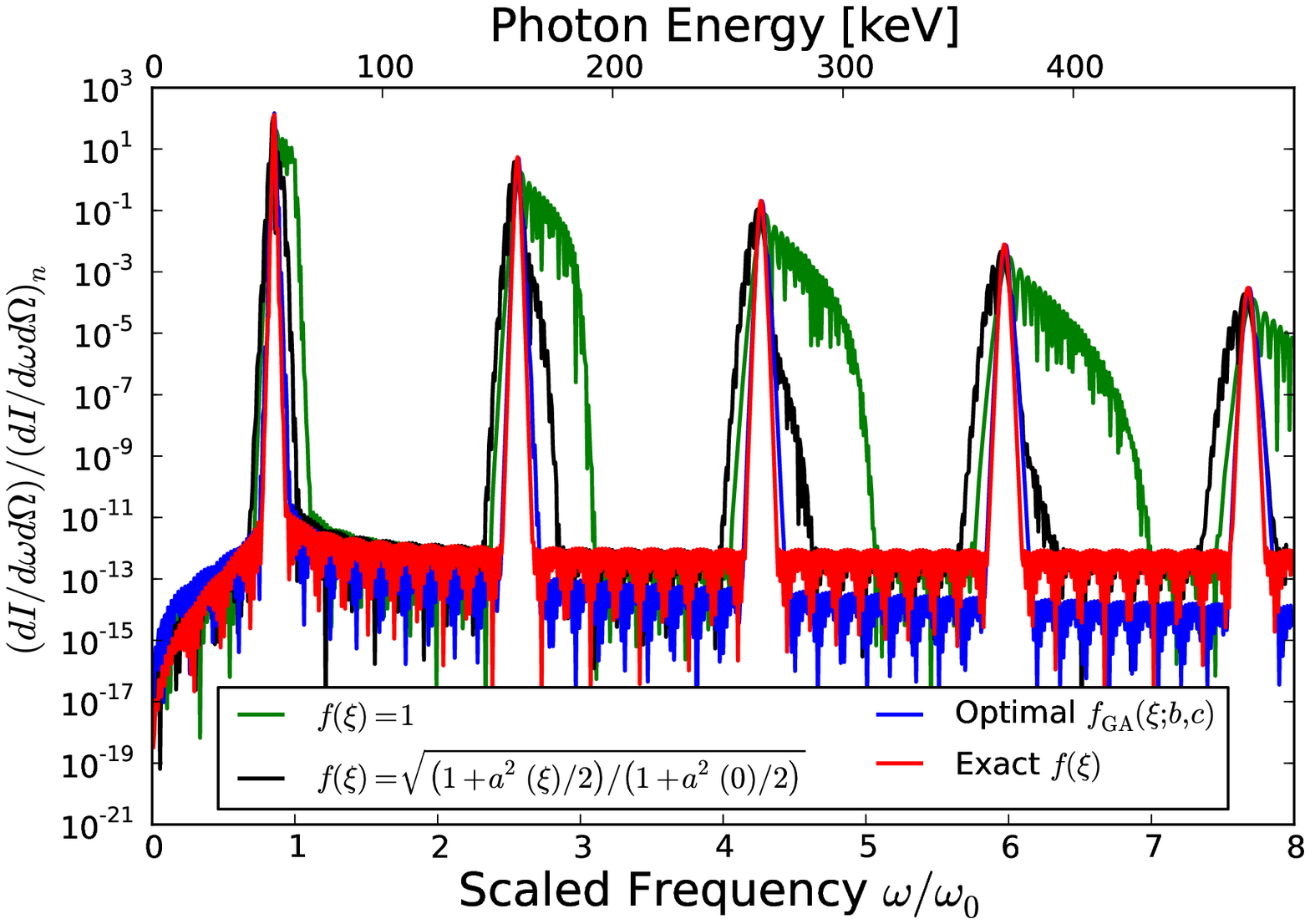} \label{fig:exp}}
\subfigure{
\includegraphics[width=3.375in]{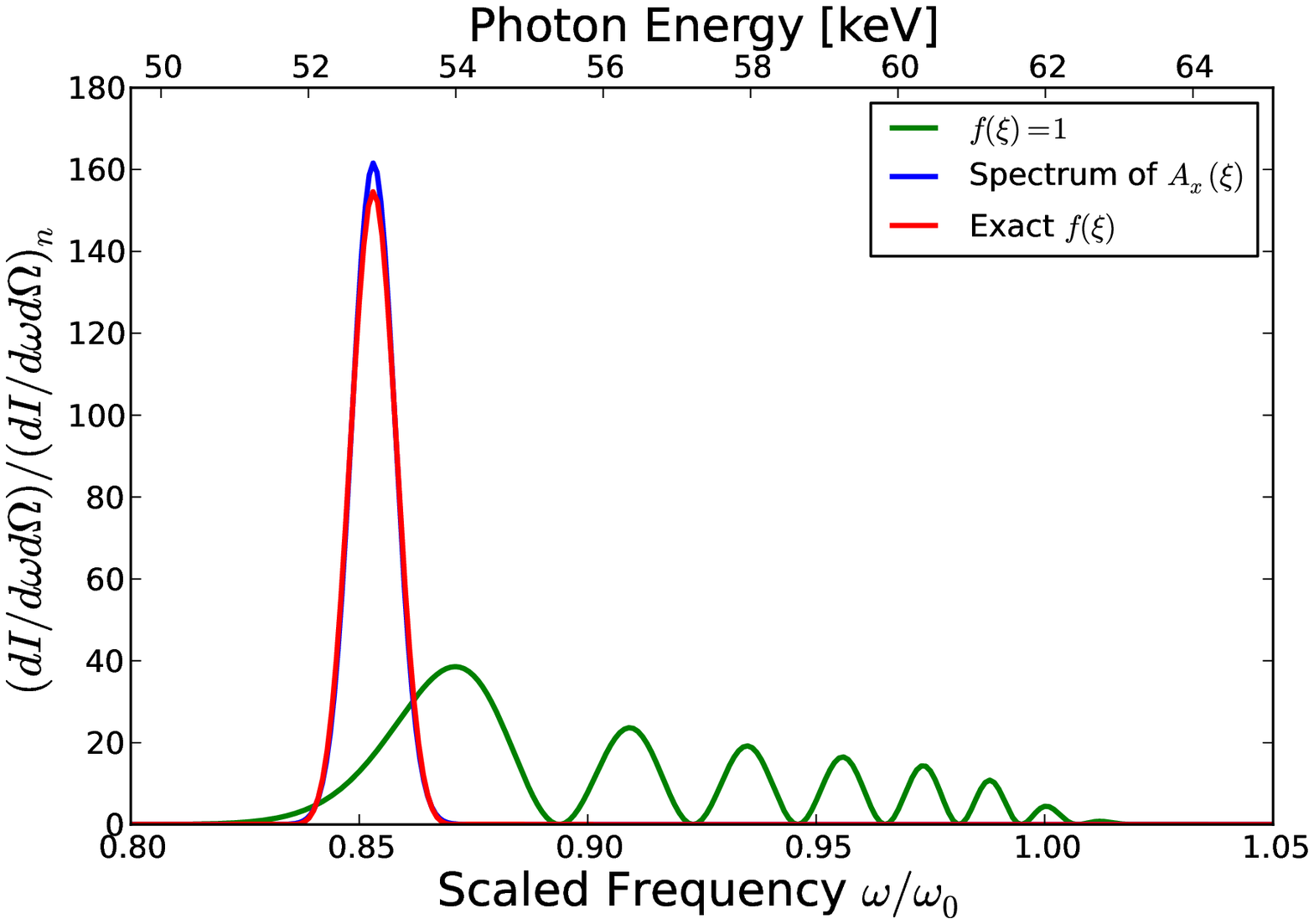} \label{fig:fft}}
\caption{\small{Top: Normalized spectra of scattered radiation for the case
without FM (green line), FM from Eq.~(\ref{eq:sqrt}) 
(black line), FM from optimal $f_{\rm GA}(\xi;b,c)$ 
(blue line), and the exact FM from Eq.~(\ref{eq:exact}) (red line).
The shapes of the spectra --- both non-modulated and modulated --- are 
independent on the scattering electron's energy. 
Bottom: Complete correction of spectral width is demonstrated by this 
comparison of the case with no FM (green line), the Fourier transform of 
the amplitude function (blue line), and the case with exact FM (red line). 
In both panels, a gaussian envelope is used with $a_0=0.587$ and 
electron's $\gamma = 100$ as in Fig.~2 of \cite{krafft}.
}}
\end{center}
\end{figure}

Our calculations were motivated by the suggestion in Ref.~\cite{gheb}. 
FM corresponding to their proposed 
$\omega(t) = 2\omega_0 \left\{1 + \left[a(t)/\sqrt{2}a_0\right]^2\right\}/3$
(found in the text below their Eq.~(4)) is
\begin{equation} \label{eq:model_23}
f(\xi)={2\over3}\left(1+{a^2(\xi)\over{2a^2(0)}}\right).
\end{equation}
As understood and discussed subsequently, this substitution is strictly 
valid only when $(\xi f(\xi))'=f(\xi)+\xi f'(\xi)\approx f(\xi)$, that is, 
the FM occurs slowly enough. We also calculated 
the spectral distributions resulting from
\begin{equation} \label{eq:sqrt}
f(\xi)=\sqrt{1+a^2(\xi)/2\over{1+a^2(0)/2}}.
\end{equation}
From Fig.~\ref{fig:exp}, it is evident that the calculated spectrum for 
Eq.~(\ref{eq:sqrt}) represents a significant improvement over the
non-modulated case. Because all harmonics improved, 
we concluded that a prescription exists to compensate all harmonics. 
Next, we investigated optimal configurations for 
$f(\xi)$ using genetic optimization
\cite{Zitzler2001b,Bleuler2002a,Bleuler2003a,bazarov,hofler}. 
The optimization simultaneously maximized the height and minimized the 
width of the fundamental peak in the spectrum for a two-parameter family 
of functions of the form $f_{\rm GA}({\bar \xi}; b, c) = 
c/{\left[1 - (1-c)\exp\left(-b{\bar \xi}^2\right)\right]}$. 
The best result has $b=0.859$ and $c=0.907$. The spectrum
of the optimal $f_{\rm GA}$, as well as the exact FM derived below is 
also shown in Fig.~\ref{fig:exp}. Genetic optimization performs 
quite well, as the spectrum of the optimal $f_{\rm GA}$ is very close to 
that of the exact FM.
%
%

We investigated the dependence of the optimal parameters of $f_{\rm GA}$
on the field strength in the incident laser pulse and observed that $c$ 
scaled directly as $a(0)^2/4$. Convinced that such simple scaling indicates 
an underlying physics cause for narrowband emission, we derived the following
analytical condition that leads to minimal emission bandwidth.

\begin{figure}
\includegraphics[width=3.375in]{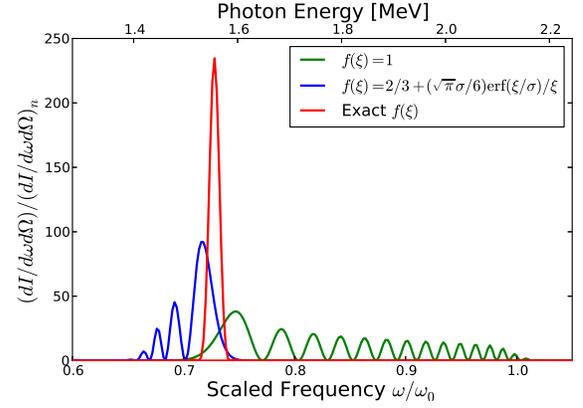}
\caption{\small{Narrowing the radiation spectra by exact FM. 
Scattered spectral distributions calculated with no FM (green line), 
the GSU FM as in Eq.~(\ref{eq:GSU}) (blue line), and the exact FM 
given in Eq.~(\ref{eq:exact_gauss}) (red line). 
A gaussian envelope is used with $a_0=0.865$ as in Fig.~6 of \cite{gheb}.
Note the linear scale and that only data in the first 
harmonic is displayed for both panels. The top $x$-axes in both panels
denote photon energies for the case of $E_{beam} = 300$ MeV, 
$\lambda = 800$ nm, FWHM pulse duration of 90 fs, as in Fig.~6 of
\cite{gheb}.}}
\label{fig:GSU}
\end{figure}

For incident laser pulses of the given form and when $f(\xi)$ is not slowly 
varying, one must define {\it local} values of the pulse frequency and 
wave number as 
\begin{equation}
\omega(\xi)= \partial\Phi/\partial t=cd\Phi/d\xi, 
\hskip15pt
k(\xi)=\partial\Phi/\partial z=d\Phi/d\xi, \nonumber
\end{equation}
where $\Phi$ is the incident wave phase. Lorentz-transforming these quantities 
into the beam frame (frame with zero average beam velocity) yields
$\omega'=(1+\beta^*)\gamma^*\omega$ and $k'=(1+\beta^*)\gamma^*k$, where
$\gamma^*(\xi)=\gamma/\sqrt{1+a(\xi)^2/2}$
and $\beta^*(\xi)=\sqrt{1-(1/\gamma^2)(1+a(\xi)^2/2)}$ 
are local values within the pulse. To minimize the spectral width 
in the lab frame one should arrange the frequency in the beam frame to 
emit radiation Doppler-shifted back to a constant frequency in the lab frame. 
Thus for all $\xi$-values we require {\it locally}
\begin{equation}
(1+\beta^*)^2\gamma^{*2}d\Phi/d\xi=C=(1+\beta)^2\gamma^2d
\Phi/d\xi(\xi=-\infty), 
\nonumber
\end{equation}
for some constant $C$. This exact case is found by integration
\begin{equation}
{d\over{d\xi}}\left[\xi f(\xi)
\right]=(1+a(\xi)^2/2)f(\xi=-\infty),
\nonumber
\end{equation}
as both $\beta$ and $\beta^*$ are close to one for relativistic scattering.
The solution with the boundary condition $f(0)=1$ is
\begin{equation} \label{eq:exact}
f(\xi)={1\over{1+a(0)^2/2}}
\left(1+{\int_0^\xi a(\xi')^2d\xi'\over2\xi}\right).
\end{equation}
Clearly, this function falls from 1 to the constant value 
$f(\pm\infty)=(1+a(0)^2/2)^{-1}$. This characteristic, that $f$ falls off 
to the same value at $\xi=\pm\infty$, is shared by any single-peaked 
symmetrical model for the amplitude function, as is clear from the general 
formula in Eq.~(\ref{eq:exact}). We can now explain the scaling in the 
optimization. The optimal $f_{\rm GA}$ closely traces the exact FM around 
$\xi=0$, while its purely exponential tails are unable to match the
exact $1/\xi$ behavior.

Applying this pulse-chirping prescription leads to narrowband emission,
as shown in Fig.~\ref{fig:exp}. The non-FM case clearly shows ponderomotive 
broadening at high field strength
that is corrected away using the exact FM prescription. Notice a non-trivial
conclusion of this calculation: our prescription works across all the 
harmonics shown, whereas Ref.~\cite{gheb} presents information on the 
fundamental line only. Additionally, because Eq.~(\ref{eq:exact}) is 
energy-independent, compensation occurs for all electrons in a bunch of
electrons with varying energy. In this case, the linewidth is spread in 
the usual way by the energy spread \cite{krafftreview}.

In Fig.~\ref{fig:fft} we compare the spectral peak in the high-field case 
with the spectrum obtained simply from the Fourier transform of the gaussian 
amplitude/envelope function $a(\xi)$, suitably shifted to the peak frequency 
of the first curve. Observe that the corrected width is identical to that
generated by the gaussian (and is Fourier-limited by the duration of 
the scattered pulse). One cannot expect to obtain a peak narrower than the 
Fourier-limited linewidth. The scattered photons in the high-field case 
have a linewidth much narrower than that of the Fourier spectrum of the 
incident frequency-modulated laser pulse.

Using this more accurate method defining the frequency and wave number, 
it is now possible to generate a scattered distribution closer to the one 
reported in Ref.~\cite{gheb}. Integrating
\begin{equation}
{d\over{d\xi}}\left[\xi f_{\rm GSU}(\xi)\right]
={2\over3}\left(1+{a(\xi)^2\over2a(0)^2}\right) \nonumber
\end{equation}
for a gaussian envelope 
yields the analogous $f_{\rm GSU}$ with the correct boundary conditions
\begin{equation} \label{eq:GSU}
f_{\rm GSU}(\xi)={2\over3}\left(1+{\sqrt{\pi}\sigma\over4\xi}
{\rm erf}(\xi/\sigma)\right).
\end{equation}
For comparison, exact FM for the gaussian envelope is
\begin{equation} \label{eq:exact_gauss}
f(\xi) = {1\over{1+a_0^2/2}} 
\left(1+{{\sqrt{\pi}\sigma a_0^2}\over{4\xi}} {\rm erf}(\xi/\sigma)\right).
\end{equation}
FM obtained using the empirical 
prescription of \cite{gheb}, and given here in Eqns.~({\ref{eq:model_23}) 
and (\ref{eq:GSU}), does not satisfy the asymptotic behavior of 
$\lim_{a(0)\to 0} f(\xi) = 1$ because they do not even depend on $a(0)$.
Therefore Eqn.~(\ref{eq:GSU}) should not be directly compared to the 
exact FM in Eq.~(\ref{eq:exact_gauss}) which satisfies this asymptotic behavior.
A cursory inspection shows that the empirical FM in
Eq.~(\ref{eq:GSU}) is equivalent to the exact FM in Eq.~(\ref{eq:exact_gauss})
for $a(0)=1$ and explains why \cite{gheb} observed a substantial
narrowing of the spectra for their case of $a(0)=0.865\approx 1$.

In Fig.~\ref{fig:GSU} we show scattered spectral distributions calculated with 
$f(\xi)=1$, the modified GSU model in Eq.~(\ref{eq:GSU}), and the fully 
corrected result using Eq.~(\ref{eq:exact_gauss}), now on a linear scale 
as in the previous publication \cite{gheb1}. We observe that pulse chirping 
increases the peak spectral energy density by a factor of 2.4 in going 
from unmodulated to the Eq.~(\ref{eq:GSU}) model, and another factor of 
2.5 in applying the exact FM prescription.

Alternatively, Eq.~(\ref{eq:exact}) follows from a 
stationary phase analysis of \cite{Brau,ck,krafft}
\begin{widetext}
\begin{eqnarray} \label{eq:Ds}
D_x & = & \int^\infty_{-\infty}{d\xi\over\gamma(1+\beta)}\tilde{A}(\xi)
\exp\left[i\omega\left({\xi\over{c\gamma^2(1+\beta)^2}}+
{1\over{c\gamma^2(1+\beta)^2}}\int_{-\infty}^\xi\tilde{A}^2(\xi')d\xi'
\right)\right] \\
& \approx & {1\over2} {\int^\infty_{-\infty}{d\xi\over\gamma(1+\beta)}a(\xi)
\exp\left[{-2\pi i \xi f(\xi)\over\lambda}
+i\omega\left({\xi\over{c\gamma^2(1+\beta)^2}}+
{1\over{c\gamma^2(1+\beta)^2}}\int_{-\infty}^\xi\tilde{A}^2(\xi')d\xi'
\right)\right]} \nonumber
\end{eqnarray}
\end{widetext}
where $d^2E/d\omega d\Omega=r_eE_{beam}\omega^2|D_x|^2/(4\pi^2 c^3\gamma)=
(d^2E/d\omega d\Omega)_n(\omega/\omega_0)^2|(1+\beta)\gamma D_x/\lambda|^2$ 
in the forward direction. The dominant contribution to $D_x$ comes from 
replacing $\tilde{A}^2(\xi)$ by its average value $a^2(\xi)/2$ in the 
phase integral. Applying Eq.~(\ref{eq:exact}) guarantees that the phase in 
the integral is constant on average and only slightly modulated as a function 
of $\xi$. Therefore, the maximum value of the 
amplitude of $D_x$ is well-approximated by 
\begin{equation}
|D_x|_{\rm max} \approx {1\over2\gamma(1+\beta)}\int_{-\infty}^\infty a(\xi)d\xi,
\nonumber
\end{equation}
as observed in the exact FM case.

The fact that the $f$ function falls from a maximum in the middle of the pulse, 
suggests an immediate practical realization of the frequency chirping. 
Suppose an FEL oscillator is constructed where the driving beam bunches are 
long enough that the RF-curvature related energy spread is substantial
(see Fig.~\ref{fig:dist}). The frequency of the resulting laser pulse 
emitted by the oscillator FEL will also be chirped: 
\begin{equation}
\omega(\xi)=\omega(\xi=0)\cos^2(2\pi\xi/\lambda_{\rm RF}), \nonumber
\end{equation}
where $\lambda_{\rm RF}$ is the wavelength of the RF accelerating wave, and 
it is assumed that the center of the electron pulse coincides with the 
accelerating wave maximum. The circulating bunches will generate a 
$\xi$-dependent frequency from the energy dependence of the FEL emission.

Using our prescription for defining $f(\xi)$,
\begin{equation}
{d\over{d\xi}}\left[\xi f_{\rm RF}(\xi)\right]=\cos^2(2\pi\xi/\lambda_{\rm RF})
\nonumber
\end{equation}
yields
\begin{equation} \label{eq:RF}
f_{\rm RF}(\xi)={1\over2}+{1\over\xi}{\lambda_{\rm RF}\over8\pi}
\sin(4\pi\xi/\lambda_{\rm RF}).
\end{equation}
Notice the dependence has the correct sign to allow compensation. 
Expanding the above equation around the center of the pulse produces
$f_{\rm RF}(\xi) \approx 1 - {{4\pi^2} \xi^2/{(3\lambda^2_{\rm RF})}}$,
and matching it to the quadratic dependence of the exact FM function in
Eq.~(\ref{eq:exact_gauss}) yields
$\lambda_{\rm RF0}=\sqrt{8}\pi{\sqrt{1+a_0^2/2}\over a_0}\sigma.$
%

\begin{table}[htbp]
\caption{\small{Peak height (normalized to peak height for exact FM) 
and optimal choice for $\lambda_{\rm RF}$ for each harmonic when approximated 
by RF waveform curvature model in Eq.~(\ref{eq:RF}).
}}
\begin{tabular}{ c  c  c  c  }
    \multicolumn{3}{c} {  } \\ \hline \hline
    Harmonic & \hskip10pt Peak Height \hskip10pt & 
    \hskip10pt $\lambda_{\rm RF}/\lambda_{\rm RF0}$ \hskip10pt \\ \hline
    1&0.84&1.47   \\ 
    3&0.89& 1.21  \\ 
    5&0.94& 1.14  \\ 
    7&0.98 & 1.10 \\ 
    9&0.99&1.12   \\ 
    \hline
    \hline
\end{tabular}
\label{tab:peak}
\end{table}

Because we cannot reproduce the optimal frequency profile exactly by these 
means, we investigated whether we could obtain peaks in the spectra of all 
reported harmonics close to those in the optimal case. Our results showing 
the spectral peak values and the optimal $\lambda_{\rm RF}$ for the various
harmonics, compared to the optimal case, are reported in Table \ref{tab:peak}.
Indeed, we are able to get close-to-optimal performance across all harmonics
with slight adjustments in RF-frequency (compressed bunch length in practice). 
Interestingly, higher harmonics achieved optimal performance by 
adjusting closer to the $\lambda_{\rm RF0}$ value.

Finally, we investigated the robustness of the solutions. Requiring the
spectrum peaks to be degraded under 10\% leads to a restriction of the 
bunch length. Because the peaks as a function of bunch 
length are quite broad, control of the bunch length at the 20\% level is 
indicated. There is already experimental evidence \cite{Zhang} that 
controlled chirping of an electron bunch driving an FEL oscillator is 
reflected in the FEL output radiation characteristics.

The cases discussed in this paper, corresponding to laser parameters of 
interest in many sources, all yield less than one photon per electron.
More quantitatively, from \cite{krafftreview} we derive the domain of 
validity of the no-electron-recoil approximation, i.e. when the number 
of emitted photons 
$n_{\gamma}=(\alpha \lambda)/(3\pi)\int_{-\infty}^{\infty} 
\left|{{\partial {\tilde A}(\xi)}/{\partial \xi}}\right|^2 
d\xi <1$, to be 
for field strength $a_0 < \sqrt{3\lambda/(2\pi^{1/2} \alpha \sigma)}$, with
$\alpha$ the fine structure constant. For the cases considered here 
this restriction
reduces to $a_0 < 2.39$, which is easily satisfied in the cases reported
in this Letter ($n_{\gamma}=0.06$ for $a_0=0.587$, $\gamma=100$ and 
$n_{\gamma}=0.13$ for $a_0=0.865$, $E_{beam}=300$ MeV case). 
Therefore, the field strength is not so large that 
quantum mechanical multi-photon emission processes need to be considered.

It is our belief that because including electron recoil in the formula for 
the emitted photon energy is straightforward, a calculation procedure similar 
to the above will lead to chirping prescriptions including the full 
Compton effect.

The calculations presented suggest the following conclusions. 
Frequency chirping of the incident laser pulse can indeed lead to bandwidth 
{\it reduction} in the radiation emerging from Thomson scattering 
events at high field strength, as suggested by GSU. This somewhat 
counterintuitive situation arises because, with proper tuning, the 
double-Doppler-shifted frequency in the lab frame may be adjusted to a 
constant value by compensating the frequency shifts due to velocity changes 
against the FM. We have analytically derived exact frequency 
compensation functions that may be applied to very general 
longitudinal pulse shapes and across all the lower-order harmonics,
which become more prominent in the high-field regime.
We have suggested a practical realization of this compensation idea in 
terms of a chirped-beam driven FEL oscillator configuration and shown 
that significant compensation can occur, even with imperfect matching. 

Discussions with S.~Benson and D.~Douglas are gratefully 
acknowledged, who assured us that FEL laser pulse chirping could be 
accomplished by electron beam chirping. S.~Benson provided references on 
early work in FEL tapering. 
R.~Ruth provided information on the work at Lyncean Technologies. 
In addition, fruitful interactions with I.~Ghebregziabher and D.~Umstadter 
are acknowledged. They graciously consented to our Fig.~\ref{fig:GSU} as 
being reported as qualitatively and quantitatively similar to their Fig.~6 
of Ref.~\cite{gheb}, even though somewhat different models were used to 
generate the two figures. Our communications with G.~P.~Williams, 
M.~Tiefenback and S.~Corneliussen were very helpful. We are thankful to
J.~Griffin for her help in generating our Fig.~\ref{fig:dist}.
This Letter is authored by Jefferson Science Associates, LLC under 
U.~S.~Department of Energy (DOE) Contract No.~DE-AC05-06OR23177.  
The U.~S.~Government retains a non-exclusive, paid-up, irrevocable, 
world-wide license to publish or reproduce this manuscript for 
U.~S.~Government purposes.
K.~D.~is supported by DOE Contract No.~DE-SC00004094.



\bibliographystyle{elsarticle-num}


\begin{thebibliography}{00}
\bibitem{gheb1} Compare with Fig.~6 of Ref.~\cite{gheb} or Fig.~3 in the archive 
version. We shared the results of our calculations with these authors, who 
have separately confirmed the qualitative and quantitative agreement between 
this model and their results.
\end{thebibliography}




\end{document}